\documentclass[preprint,showpacs,preprintnumbers]{revtex4-2}
\usepackage{amsmath,amssymb}
\usepackage{graphicx}
\usepackage{xcolor}
\usepackage{url}
\usepackage[font={scriptsize,stretch=1.0},labelfont={bf,it},textfont=it]{caption}

\begin{document}

\def\bx{{\bf x}}
\def\bxa{{\bf x}_A}
\def\bxb{{\bf x}_B}
\def\bxkl{{\bf x}_{k,l}}
\def\bxk{{\bf x}_{j}}
\def\bxbj{{\bf x}_B^{(k)}}
\def\di{\partial_i}
\def\d3{\partial_3}
\def\dt{\partial_t}

\newcommand{\dopB}[1]{\mathfrak{D}_B^\theta\{#1\}}
\newcommand{\dopBk}[1]{\mathfrak{D}_B^{\theta,(k)}\{#1\}}

\def\dD{\partial\mathbb{D}}
\def\dDo{\partial\mathbb{D}_0}
\def\dDa{\partial\mathbb{D}_A}

\def \gxklxaw {G(\bxkl,\bxa,\omega)}
\def \gxaxklw {G(\bxa,\bxkl,\omega)}
\def \gxklxamw {G^*(\bxkl,\bxa,\omega)}
\def \gxaxklmw {G^*(\bxa,\bxkl,\omega)}
\def \gxklxbw {G(\bxkl,\bxb,\omega)}
\def \gxbxklw {G(\bxb,\bxkl,\omega)}
\def \gxklxbmw {G^*(\bxkl,\bxb,\omega)}
\def \gxbxklmw {G (\bxb,\bxkl,\omega)}

\def \pxklxaw {p(\bxkl,\bxa,\omega)}
\def \pxaxklw {p(\bxa,\bxkl,\omega)}
\def \pxklxamw {p^*(\bxkl,\bxa,\omega)}
\def \pxaxklmw {p^*(\bxa,\bxkl,\omega)}
\def \pxklxbw {p(\bxkl,\bxb,\omega)}
\def \pxbxklw {p(\bxb,\bxkl,\omega)}
\def \pxklxbmw {p^*(\bxkl,\bxb,\omega)}
\def \pxbxklmw {p (\bxb,\bxkl,\omega)}

\def \gxkxaw {G(\bxk,\bxa,\omega)}
\def \gxaxkw {G(\bxa,\bxk,\omega)}
\def \gxkxamw {G^*(\bxk,\bxa,\omega)}
\def \gxaxkmw {G^*(\bxa,\bxk,\omega)}
\def \gxkxbw {G(\bxk,\bxb,\omega)}
\def \gxbxkw {G(\bxb,\bxk,\omega)}
\def \gxkxbmw {G^*(\bxk,\bxb,\omega)}
\def \gxbxkmw {G (\bxb,\bxk,\omega)}

\def \pxkxaw {p(\bxk,\bxa,\omega)}
\def \pxaxkw {p(\bxa,\bxk,\omega)}
\def \pxkxamw {p^*(\bxk,\bxa,\omega)}
\def \pxaxkmw {p^*(\bxa,\bxk,\omega)}
\def \pxkxbw {p(\bxk,\bxb,\omega)}
\def \pxbxkw {p(\bxb,\bxk,\omega)}
\def \pxkxbmw {p^*(\bxk,\bxb,\omega)}
\def \pxbxkmw {p (\bxb,\bxk,\omega)}

\def \pxkxbjw {p(\bxk,\bxbj,\omega)}
\def \pxbjxkw {p(\bxbj,\bxk,\omega)}
\def \pxkxbjmw {p^*(\bxk,\bxbj,\omega)}
\def \pxbjxkmw {p (\bxbj,\bxk,\omega)}
\def \pxbxkjmw {p^*(\bxbj,\bxk,\omega)}
\def \pxkjxbkmw {p (\bxk,\bxbj,\omega)}

\def \gxxat {G(\bx,\bxa,t)}
\def \gxaxt {G(\bxa,\bx,t)}
\def \gxxamt {G(\bx,\bxa,-t)}
\def \gxaxmt {G(\bxa,\bx,-t)}
\def \gxxbt {G(\bx,\bxb,t)}
\def \gxbxt {G(\bxb,\bx,t)}
\def \gxxbmt {G(\bx,\bxb,-t)}
\def \gxbxmt {G(\bxb,\bx,-t)}
\def \ghxxat {G_{\rm h}(\bx,\bxa,t)}
\def \ghxaxt {G_{\rm h}(\bxa,\bx,t)}
\def \gxbxat {G(\bxb,\bxa,t)}
\def \gxaxbt {G(\bxa,\bxb,t)}
\def \ghxbxat {G_{\rm h}(\bxb,\bxa,t)}
\def \ghxaxbt {G_{\rm h}(\bxa,\bxb,t)}

\def \gxxaw {G(\bx,\bxa,\omega)}
\def \gxaxw {G(\bxa,\bx,\omega)}
\def \gxxamw {G^*(\bx,\bxa,\omega)}
\def \gxaxmw {G^*(\bxa,\bx,\omega)}
\def \gxxbw {G(\bx,\bxb,\omega)}
\def \gxbxw {G(\bxb,\bx,\omega)}
\def \gxxbmw {G^*(\bx,\bxb,\omega)}
\def \gxbxmw {G^*(\bxb,\bx,\omega)}
\def \gxbxaw {G(\bxb,\bxa,\omega)}
\def \gxaxbw {G(\bxa,\bxb,\omega)}
\def \ghxxaw {G_{\rm h}(\bx,\bxa,\omega)}
\def \ghxaxw {G_{\rm h}(\bxa,\bx,\omega)}
\def \ghxbxaw {G_{\rm h}(\bxb,\bxa,\omega)}
\def \ghxaxbw {G_{\rm h}(\bxa,\bxb,\omega)}

\def \gxxbjt {G(\bx,\bxbj,t)}
\def \gxbjxt {G(\bxbj,\bx,t)}
\def \gxxbjmt {G(\bx,\bxbj,-t)}
\def \gxbjxmt {G(\bxbj,\bx,-t)}
\def \gxbjxat {G(\bxbj,\bxa,t)}
\def \gxaxbjt {G(\bxa,\bxbj,t)}
\def \ghxbjxat {G_{\rm h}(\bxbj,\bxa,t)}
\def \ghxaxbjt {G_{\rm h}(\bxa,\bxbj,t)}

\def \gxxbjw {G(\bx,\bxbj,\omega)}
\def \gxbjxw {G(\bxbj,\bx,\omega)}
\def \gxxbjmw {G^*(\bx,\bxbj,\omega)}
\def \gxbjxmw {G^*(\bxbj,\bx,\omega)}
\def \gxbjxaw {G(\bxbj,\bxa,\omega)}
\def \gxaxbjw {G(\bxa,\bxbj,\omega)}
\def \ghxbjxaw {G_{\rm h}(\bxbj,\bxa,\omega)}
\def \ghxaxbjw {G_{\rm h}(\bxa,\bxbj,\omega)}

\def \pxxat {p(\bx,\bxa,t)}
\def \pxaxt {p(\bxa,\bx,t)}
\def \pxxamt {p(\bx,\bxa,-t)}
\def \pxaxmt {p(\bxa,\bx,-t)}
\def \pxxbt {p(\bx,\bxb,t)}
\def \pxkxbt {p(\bxk,\bxb,t)}
\def \pxbxt {p(\bxb,\bx,t)}
\def \pxxbmt {p(\bx,\bxb,-t)}
\def \pxbxmt {p(\bxb,\bx,-t)}
\def \phxxat {p_{\rm h}(\bx,\bxa,t)}
\def \phxaxt {p_{\rm h}(\bxa,\bx,t)}
\def \phxbxat {p_{\rm h}(\bxb,\bxa,t)}
\def \phxaxbt {p_{\rm h}(\bxa,\bxb,t)}
\def \pxbxat {p(\bxb,\bxa,t)}
\def \pxaxbt {p(\bxa,\bxb,t)}
\def \pxbxamt {p(\bxb,\bxa,-t)}
\def \pxaxbmt {p(\bxa,\bxb,-t)}

\def \pxxaw {p(\bx,\bxa,\omega)}
\def \pxaxw {p(\bxa,\bx,\omega)}
\def \pxxamw {p^*(\bx,\bxa,\omega)}
\def \pxaxmw {p^*(\bxa,\bx,\omega)}
\def \pxxbw {p(\bx,\bxb,\omega)}
\def \pxbxw {p(\bxb,\bx,\omega)}
\def \pxaxbw {p(\bxa,\bxb,\omega)}
\def \pxbxaw {p(\bxb,\bxa,\omega)}
\def \pxxbmw {p^*(\bx,\bxb,\omega)}
\def \pxbxmw {p^*(\bxb,\bx,\omega)}
\def \phxxaw {p_{\rm h}(\bx,\bxa,\omega)}
\def \phxaxw {p_{\rm h}(\bxa,\bx,\omega)}
\def \phxbxaw {p_{\rm h}(\bxb,\bxa,\omega)}
\def \phxaxbw {p_{\rm h}(\bxa,\bxb,\omega)}
\def \pxbxaw {p(\bxb,\bxa,\omega)}
\def \pxaxbw {p(\bxa,\bxb,\omega)}
\def \pxbxamw {p^*(\bxb,\bxa,\omega)}
\def \pxaxbmw {p^*(\bxa,\bxb,\omega)}

\def \pxxbjt {p(\bx,\bxbj,t)}
\def \pxbjxt {p(\bxbj,\bx,t)}
\def \pxxbjmt {p(\bx,\bxbj,-t)}
\def \pxbjxmt {p(\bxbj,\bx,-t)}
\def \phxbjxat {p_{\rm h}(\bxbj,\bxa,t)}
\def \phxaxbjt {p_{\rm h}(\bxa,\bxbj,t)}
\def \pxbjxat {p(\bxbj,\bxa,t)}
\def \pxaxbjt {p(\bxa,\bxbj,t)}
\def \pxbjxamt {p(\bxbj,\bxa,-t)}
\def \pxaxbjmt {p(\bxa,\bxbj,-t)}

\def \phxaxbjtt {p_{\rm h}(\bxa,\bxbj,t-t^{(k)})}

\def \pxxbjw {p(\bx,\bxbj,\omega)}
\def \pxbjxw {p(\bxbj,\bx,\omega)}
\def \pxaxbjw {p(\bxa,\bxbj,\omega)}
\def \pxbjxaw {p(\bxbj,\bxa,\omega)}
\def \pxxbjmw {p^*(\bx,\bxbj,\omega)}
\def \pxbjxmw {p^*(\bxbj,\bx,\omega)}
\def \phxbjxaw {p_{\rm h}(\bxbj,\bxa,\omega)}
\def \phxaxbjw {p_{\rm h}(\bxa,\bxbj,\omega)}
\def \pxbjxaw {p(\bxbj,\bxa,\omega)}
\def \pxaxbjw {p(\bxa,\bxbj,\omega)}
\def \pxbjxamw {p^*(\bxbj,\bxa,\omega)}
\def \pxaxbjmw {p^*(\bxa,\bxbj,\omega)}

\def \fxxat {f_1(\bx,\bxa,t)}
\def \fxaxt {f_1(\bxa,\bx,t)}
\def \fxxbt {f_1(\bx,\bxb,t)}
\def \fxbxt {f_1(\bxb,\bx,t)}
\def \fxaxbt {f_1(\bxa,\bxb,t)}
\def \fxbxat {f_1(\bxb,\bxa,t)}

\def \fxxaw {f_1(\bx,\bxa,\omega)}
\def \fxaxw {f_1(\bxa,\bx,\omega)}
\def \fxxbw {f_1(\bx,\bxb,\omega)}
\def \fxbxw {f_1(\bxb,\bx,\omega)}
\def \fxaxbw {f_1(\bxa,\bxb,\omega)}
\def \fxbxaw {f_1(\bxb,\bxa,\omega)}

\def \fxklxaw {f_1(\bx,\bxa,\omega)}
\def \fxaxklw {f_1(\bxa,\bxkl,\omega)}

\def \fpxxat {f^+_1(\bx,\bxa,t)}
\def \fpxaxt {f^+_1(\bxa,\bx,t)}
\def \fpxxbt {f^+_1(\bx,\bxb,t)}
\def \fpxbxt {f^+_1(\bxb,\bx,t)}
\def \fpxaxbt {f^+_1(\bxa,\bxb,t)}
\def \fpxbxat {f^+_1(\bxb,\bxa,t)}

\def \fpxxaw {f^+_1(\bx,\bxa,\omega)}
\def \fpxaxw {f^+_1(\bxa,\bx,\omega)}
\def \fpxxbw {f^+_1(\bx,\bxb,\omega)}
\def \fpxbxw {f^+_1(\bxb,\bx,\omega)}
\def \fpxaxbw {f^+_1(\bxa,\bxb,\omega)}
\def \fpxbxaw {f^+_1(\bxb,\bxa,\omega)}

\def \fpxxbjw {f^+_1(\bx,\bxbj,\omega)}
\def \fpxbjxw {f^+_1(\bxbj,\bx,\omega)}
\def \fpxaxbjw {f^+_1(\bxa,\bxbj,\omega)}
\def \fpxbjxaw {f^+_1(\bxbj,\bxa,\omega)}
\def \fxaxbjw {f_1(\bxa,\bxbj,\omega)}
\def \fxbjxaw {f_1(\bxbj,\bxa,\omega)}

\def \fmxxat {f^-_1(\bx,\bxa,t)}
\def \fmxaxt {f^-_1(\bxa,\bx,t)}
\def \fmxxbt {f^-_1(\bx,\bxb,t)}
\def \fmxbxt {f^-_1(\bxb,\bx,t)}
\def \fmxaxbt {f^-_1(\bxa,\bxb,t)}
\def \fmxbxat {f^-_1(\bxb,\bxa,t)}

\def \fmxxaw {f^-_1(\bx,\bxa,\omega)}
\def \fmxaxw {f^-_1(\bxa,\bx,\omega)}
\def \fmxxbw {f^-_1(\bx,\bxb,\omega)}
\def \fmxbxw {f^-_1(\bxb,\bx,\omega)}
\def \fmxaxbw {f^-_1(\bxa,\bxb,\omega)}
\def \fmxbxaw {f^-_1(\bxb,\bxa,\omega)}

\def \fmxxbjw {f^-_1(\bx,\bxbj,\omega)}
\def \fmxbjxw {f^-_1(\bxbj,\bx,\omega)}
\def \fmxaxbjw {f^-_1(\bxa,\bxbj,\omega)}
\def \fmxbjxaw {f^-_1(\bxbj,\bxa,\omega)}

\def \fpxklxaw {f^+_1(\bxkl,\bxa,\omega)}
\def \fpxaxklw {f^+_1(\bxa,\bxkl,\omega)}
\def \fmxklxaw {f^-_1(\bxkl,\bxa,\omega)}
\def \fmxaxklw {f^-_1(\bxa,\bxkl,\omega)}

\def \fpxkxaw {f^+_1(\bxk,\bxa,\omega)}
\def \fpxaxkw {f^+_1(\bxa,\bxk,\omega)}
\def \fmxkxaw {f^-_1(\bxk,\bxa,\omega)}
\def \fmxaxkw {f^-_1(\bxa,\bxk,\omega)}
\def \fxkxaw {f_1(\bxk,\bxa,\omega)}
\def \fxaxkw {f_1(\bxa,\bxk,\omega)}

\def \Pxat {P(\bxa,t)}
\def \Pxamt {P(\bxa,t)}
\def \Pxaw {P(\bxa,\omega)}
\def \Pxamw {P^*(\bxa,\omega)}

\def \Pxt {P(\bx,t)}
\def \Pxmt {P(\bx,t)}
\def \Pxw {P(\bx,\omega)}
\def \Pxmw {P^*(\bx,\omega)}

\def \Pxkt {P(\bxk,t)}
\def \Pxkmt {P(\bxk,t)}
\def \Pxkw {P(\bxk,\omega)}
\def \Pxkmw {P^*(\bxk,\omega)}

\def \sin {{\rm sin}}
\def \cos {{\rm cos}}
\def \di {\partial_i}

\def \pap {p_A^+}
\def \pam {p_A^-}
\def \pbp {p_B^+}
\def \pbm {p_B^-}
\def \dpap {\d3 p_A^+}
\def \dpam {\d3 p_A^-}
\def \dpbp {\d3 p_B^+}
\def \dpbm {\d3 p_B^-}
\def \qap {q_A^+}
\def \qam {q_A^-}
\def \qbp {q_B^+}
\def \qbm {q_B^-}

\def \papm {p_A^{+,*}}
\def \pamm {p_A^{-,*}}
\def \pbpm {p_B^{+,*}}
\def \pbmm {p_B^{-,*}}
\def \dpapm {\d3 p_A^{+,*}}
\def \dpamm {\d3 p_A^{-,*}}
\def \dpbpm {\d3 p_B^{+,*}}
\def \dpbmm {\d3 p_B^{-,*}}
\def \qapm {q_A^{+,*}}
\def \qamm {q_A^{-,*}}
\def \qbpm {q_B^{+,*}}
\def \qbmm {q_B^{-,*}}

\newcommand{\eqnref}[1]{Equation (\ref{#1})}
\newcommand{\figref}[1]{Figure \ref{#1}}
\newcommand{\eqnsref}[1]{Equations (\ref{#1})}
\newcommand{\figsref}[1]{Figures \ref{#1}}

\title{3D virtual seismology}

\author{\textbf{Joeri~Brackenhoff,~Jan~Thorbecke~and~Kees~Wapenaar}
\\
\thanks{J. Brackenhoff performed this research at the Department of Geoscience and Engineering, Delft University of Technology, Delft, the Netherlands and is currently with the Institute of Geophysics of ETH Z\"urich in Switzerland, e-mail: johannes.brackenhoff@erdw.ethz.ch. J. Thorbecke and K. Wapenaar are both with the Department of Geoscience and Engineering, Delft University of Technology, Delft, the Netherlands, e-mail: J.W.Thorbecke@tudelft.nl and C.P.A.Wapenaar@tudelft.nl.}}%

\begin{abstract}
We create virtual sources and receivers in a 3D subsurface using the previously derived single-sided homogeneous Green's function representation. We employ Green's functions and focusing functions that are obtained using reflection data at the Earth's surface, a macro velocity model and the Marchenko method. The homogeneous Green's function is a Green's function superposed with its time-reversal. Unlike the classical homogeneous Green's function representation, our approach requires no receivers on an enclosing boundary, however, it does require the source signal to be symmetric in time. We demonstrate that in 3D, the single-sided representation is an improvement over the classical representation by applying the representations to numerical data. We retrieve responses to virtual point sources with an isotropic and with a double-couple radiation pattern and compare the results to a directly modeled reference result. We also demonstrate the application of the single-sided representation for retrieving the response to a virtual rupture that consists of a superposition of double-couple point sources. This is achieved by obtaining the homogeneous Green's function for each source separately, before they are transformed to the causal Green's function, time-shifted and superposed. The single-sided representation is also used to monitor the complete wavefield that is caused by a numerically modeled rupture. However, the source signal of an actual rupture is not symmetric in time and the single-sided represenation can therefore only be used to obtain the causal Green's function. This approach leaves artifacts in the final result, however, these artifacts are limited in space and time.
\end{abstract}

\maketitle

\section{Introduction}
Over the past few decades, the amount of induced seismicity has increased and is occurring at locations around the world \cite{keranen2018induced}. While the effects of induced seismicity are often harmful, the measurements of these events can be used to gain more insight into the mechanics of earthquake rupture \cite{galis2017induced}. For example, the measurements can be used in an inversion process to obtain the seismic moment tensor, which describes the source mechanism of a seismic event \cite{aki2002quantitative}. The knowledge of the moment tensor as well as the location of the source can help to determine what caused the induced seismicity. These inversions often rely on an accurate velocity model of the subsurface to obtain the required wavefields \cite{willacy2018application}, because errors in the velocity model can cause mistakes in the inversion result \cite{vsileny2009resolution}.

A recent development for obtaining accurate wavefields in the subsurface is the homogeneous Green's function retrieval method. A homogeneous Green's function is a Green's function superposed with its time-reversal. The classical representation of the homogeneous Green's function was used for optical holography \cite{porter1970diffraction}, inverse source problems \cite{porter1982holography}, inverse scattering methods \cite{oristaglio1989inverse}, seismic imaging \cite{Esmersoy88GEO} and seismic holography \cite{Lindsey2004AJSS}. The classical representation of the homogeneous Green's function involves an integral over a closed boundary. In practical situations,  data are usually available only on an open boundary. Methods like seismic imaging and holography still work well for this situation as long as only primary waves are considered. However, internal multiples are incorrectly handled and lead to artifacts when the classical representation is approximated by an integral along an open boundary. 

Instead of the classical representation of the homogeneous Green's function, a single-sided representation can be used, which is designed to work with an open boundary, typically the surface of the Earth \cite{wapenaar2016singleGH}. This single-sided representation is designed to correctly handle the internal multiples by employing so-called focusing functions. These focusing functions can be obtained through the use of the Marchenko method, which employs reflection data at the surface of the Earth \cite{lomas2019introduction}. The single-sided representation has been successfully applied to field data \cite{brackenhoff2019virtual}.

While many applications of the Marchenko method have been performed on 2D data, recently more applications on 3D data have been achieved \cite{pereira2019internal,staring2019interbed}. Especially in areas where there are strong out-of-plane effects, the 2D approximation on 3D data can cause errors in the result \cite{lomas2020marchenko}. To properly take into account the effects of wave propagation and scattering in 3D, the single-sided retrieval scheme for the homogeneous Green's function needs to be employed together with a 3D version of the Marchenko method.

In this paper, we present the retrieval of the homogeneous Green's function in the subsurface, similar to how the 2D homogeneous Green's function was previously retrieved \cite{brackenhoff2019}, but extended to 3D. We first review the classical and single-sided homogeneous Green's function retrieval schemes and apply the schemes to single source-receiver pairs. We use a 3D Opensource Marchenko method on a synthetic reflection response, that was modeled using a subset of the Overthrust model \cite{aminzadeh1997}, to create the required Green's functions and focusing functions for the retrieval schemes. We demonstrate the method for point sources that have an isotropic radiation pattern and compare the retrieved Green's functions to directly modeled data. Furthermore, we also retrieve snapshots of wavefields at virtual receivers in 3D to observe the propagation of the wavefield through the medium over time. Aside from considering an isotropic radiation pattern, we also consider the non-isotropic double-couple radiation pattern, which describes the seismic response to a pure shear fault \cite{aki2002quantitative}. Furthermore, we consider the 3D retrieval of a response caused by a rupture in the subsurface by employing a series of superposed point sources with varying amplitudes and activation times and a double-couple radiation pattern, similar to previous research on 2D data \cite{brackenhoff2019}. For this latter situation we use two different approaches. One is a one-step process, where we assume that we measure the response from the rupture directly, so that we can monitor the wavefield as it propagates through the subsurface. Hence, in this one-step process we create virtual receivers to monitor the response to a real source. The other is a two-step process, where we use the Marchenko method to obtain the homogeneous Green's function for each virtual source point separately, and superpose them after each homogeneous Green's function has been obtained. Hence, in this two-step process we create virtual receivers and virtual sources. This is a way to forecast the wavefield that would be caused by the rupture, given the properties of the rupture and reflection data at the surface. We illustrate the methods with numerical examples. When we speak, for the sake of argument, of measurements of the response to a real source, in the examples these measurements are simulated by numerical modeling.

\section{3D Virtual Seismology}
\subsection{Wavefields}
We consider a Green's function, $G=\gxxat$, which describes the response of a medium at time $t$ and position $\bx=(x_1,x_2,x_3),$ due to an impulsive point source at $\bxa$, using a Cartesian coordinate system. In the coordinate system that we use, the third principal direction points downwards. The Green's function is the solution to the following acoustic wave equation:
\begin{equation} \label{waveq}
\di(\rho^{-1}\di G)-\kappa\dt^2 G=-\delta(\bx-\bxa)\dt\delta(t),
\end{equation}
where $\rho=\rho(\bx)$ is the density of the medium in {kg m$^{-3}$}, $\kappa=\kappa(\bx)$ is the compressibility in {kg$^{-1}$ m s$^2$}, $\di=\partial/\partial x_i$ is the component of a vector consisting of the partial differential operators in the three principal directions of the coordinate system, $\dt=\partial/\partial t$ is the temporal partial differential operator and $\delta(\cdot)$ is a Dirac delta function. In case of repeating subscripts, Einstein's summation convention applies. The Green's function is causal; i.e. $\gxxat=0$ for $t<0$, hence, it propagates away from the source location; and obeys source-receiver reciprocity so that $\gxxat=\gxaxt$. Because the wave equation for the Green's function contains a temporal derivative in the source term, the source is defined as a volume injection rate source. 

We also consider the homogeneous Green's function $G_{\rm h}=\ghxxat$, which is defined as
\begin{equation} \label{ghdef}
\ghxxat=\gxxat+\gxxamt,
\end{equation}
where $\gxxamt$ is the time-reversed Green's function, which is acausal; i.e. $\gxxamt=0$ for $t>0$, hence, it propagates towards the source. By combining \eqnsref{waveq} and (\ref{ghdef}), we obtain the acoustic homogeneous wave equation
\begin{equation} \label{hwaveq}
\di(\rho^{-1}\di G_{\rm h})-\kappa\dt^2 G_{\rm h}=0,
\end{equation}
where the right hand side vanishes, because the source term on the right hand side of \eqnref{waveq} contains a temporal derivative, hence, the wave equation for the time reversal of the Green's function causes the source term to obtain the opposite sign. 

The data that are considered in this paper are band-limited and therefore we define a pressure wavefield $\pxxat$, which is related to the Green's function by
\begin{equation} \label{pdeft}
  \pxxat=\int_{-\infty}^{\infty}G(\bx,\bxa,t-t') s(t'){\rm d}t',
\end{equation}    
where $s(t)$ indicates a specific source signal. We also define a homogeneous pressure wavefield, similar to \eqnref{ghdef},
\begin{equation} \label{phdeft}
  \phxxat=\int_{-\infty}^{\infty}G_{\rm h}(\bx,\bxa,t-t') s(t'){\rm d}t'.
\end{equation}    
Note that in \eqnref{phdeft}, we have defined that homogeneous wavefield as the convolution of the source wavelet with the homogeneous Green's function; i.e. the Green's function is superposed with its time-reversal before the convolution. If the Green's function is convolved with a wavelet before the time-reversal and the superposition is applied, the time-reversal will affect the source wavelet as well. Only when $s(t)=s(-t)$ can the convolution be applied before the superposition.

\subsection{Homogeneous Green's function retrieval}
Homogeneous Green's function retrieval has been employed in the past to obtain the response between two locations inside a medium. The classical representation states that the response between a source and receiver inside a lossless medium can be obtained if observations are available on a closed boundary $\dD$ around the medium $\mathbb{D}$ \cite{porter1970diffraction,porter1982holography,oristaglio1989inverse}. It can be written as
\begin{equation} \label{gretclass}
\begin{split}
\ghxaxbw&\\
=\oint_{\dD}\frac{-1}{i\omega\rho(\bx)}\Big\{ &\di\gxaxmw\gxxbw \\
-&\gxaxmw\di\gxxbw\Big\}n_i{\rm d}^2\bx,
\end{split}
\end{equation}
where $\ghxaxbw$ is the frequency domain version of $\ghxaxbt$, $\omega$ is the angular frequency and $n_i$ is the $i$th component of the outward pointing normal vector on $\dD$. Note that we use $\exp(i\omega t)$ in the forward Fourier transform and $\exp(-i\omega t)$ in the inverse Fourier transform. In \eqnref{gretclass}, $\gxxbw$ describes the response to a source at $\bxb$, inside the medium in $\mathbb{D}$, observed at location $\bx$ on $\dD$. $\gxaxmw$ back propagates this response from location $\bx$ at the boundary to receiver location $\bxa$ inside $\mathbb{D}$. This creates the response $\ghxaxbw$, with a source at location $\bxb$ and a receiver at location $\bxa$. \eqnref{gretclass} can be simplified by assuming that $\dD$ is a smooth boundary and the medium outside $\mathbb{D}$ is homogeneous. In this case the terms on the right hand side of \eqnref{gretclass} are nearly identical, however, they have opposite signs, resulting in
\begin{equation} \label{gretclass2}
\begin{split}
\ghxaxbw&\\
=\oint_{\dD}\frac{-2}{i\omega\rho_0}&\Big\{ \di\gxaxmw\gxxbw\Big\}n_i{\rm d}^2\bx,
\end{split}
\end{equation}
where $\rho_0$ is the density at the boundary $\dD$. The main practical disadvantage of \eqnsref{gretclass} and \eqref{gretclass2} is that a closed boundary around the medium is required, which is usually not feasible for seismological applications. More realistically, the boundary will be open and situated on a single side of the medium, which is often the surface of the Earth. In this case, the representation is approximated as
\begin{equation} \label{gretclassapp}
\begin{split}
\ghxaxbw&\\
\approx\int_{\dDo}\frac{2}{i\omega\rho_0}\Big\{ &\d3\gxaxmw\gxxbw\Big\}{\rm d}^2\bx,
\end{split}
\end{equation}
where $\dDo$ is a horizontal single open boundary and we used ${\bf n}=(0,0,-1)$. Note that we assume that the medium above $\dDo$ is homogeneous. Applying the representation in this way introduces significant artifacts in the homogeneous Green's function \cite{brackenhoff2019virtual}.

In more recent years, the homogeneous Green's function representation has been adjusted to take into account the single-sided open boundary \cite{wapenaar2016singleGH}. The scheme that is used in this paper is taken from previous research \cite[Equations (10) and (11)]{brackenhoff2019},
\begin{equation} \label{gret}
\begin{split}
\gxaxbw+&\chi(\bxb)2i\Im\{\fxbxaw\}\\
=\int_{\dDo}\frac{2}{i\omega\rho_0}&\gxxbw\\
\times\d3\Big(&\fpxxaw-\{\fmxxaw\}^*\Big){\rm d}^2\bx,
\end{split}
\end{equation}
where $\fpxbxaw$ and $\fmxbxaw$ are the decomposed focusing function (discussed below), $\fxbxaw=\fpxbxaw+\fmxbxaw$, $\Im$ denotes the imaginary part of a complex function and $\chi(\bxb)$ is a characteristic function that is defined as 
\begin{equation} \label{chixb}
\chi(\bxb)=
\begin{cases}
1,& \text{ for $\bxb$ in $\mathbb{D}$},\\
\frac{1}{2},& \text{ for $\bxb$ on $\dD=\dDo\cup\dDa$},\\
0,& \text{ for $\bxb$ outside $\mathbb{D}\cup\dD$},
\end{cases}
\end{equation}
where $\dDa$ is a horizontal open boundary inside the subsurface of the Earth at the same depth as $\bxa$. The medium in $\mathbb{D}$ is assumed to be lossless and evanescent waves are ignored. Note, that with \eqnref{gret}, we retrieve the causal Green's function instead of the homogeneous Green's function. 

In this single-sided representation, no time-reversed Green's function is employed, but rather the decomposed focusing functions $\fpxxaw$ and $\fmxxaw$ are used, where the superscripts $+$ and $-$ indicate a downgoing and upgoing wavefield, respectively, at $\bx$. These focusing functions are designed to focus from a single-sided open boundary $\dDo$ to a location $\bxa$ inside the subsurface of the Earth, generally referred to as the focal location, without artifacts caused by multiple scattering in the overburden. To achieve this, the downgoing focusing function $\fpxxaw$ for $\bx$ at $\dDo$ is defined as the inverse of a modified transmission response of the truncated medium between $\dDo$ and $\dDa$, see Appendix A of \cite{wapenaar2014marchenko} for details. The upgoing focusing function $\fmxxaw$ for $\bx$ at $\dDo$ is defined as the reflection response of the truncated medium to $\fpxxaw$.

In \eqnref{gret}, the focusing functions $\fpxxaw$ and $\fmxxaw$ operate in a similar way as the time-reversed Green's function $\gxaxmw$ does in \eqnref{gretclassapp}, backpropagating the response from the boundary $\dDo$ to location $\bxa$. The main difference is that unlike \eqnref{gretclassapp}, \eqnref{gret} is specifically designed for application to the open boundary.

The representation in \eqnref{gret} does have an issue on the left hand side of the equation in the form of the term $\chi(\bxb)2i\Im\{\fxbxaw\}$. Depending on the relative locations of the virtual receiver $\bxa$ and the virtual source $\bxb$, as formulated by \eqnref{chixb}, artifacts related to the focusing function between the two locations are introduced in the obtained Green's function. When the receiver is located above the source, the Green's function is retrieved without artifacts. When the virtual source is located at the same depth level or above the virtual receiver, artifacts are present in the retrieved Green's function. By combining \eqnsref{gret} and the Fourier transform of \eqref{ghdef}, we obtain the single-sided retrieval scheme for the homogeneous Green's function \cite[Equation (33)]{wapenaar2019green}:
\begin{equation} \label{ghret}
\begin{split}
\ghxaxbw&=4\Re\int_{\dDo}\frac{1}{i\omega\rho_0}\gxxbw\\
\times\d3\Big(&\fpxxaw-\{\fmxxaw\}^*\Big){\rm d}^2\bx.
\end{split}
\end{equation}
\eqnref{ghret} expresses the retrieval of the homogeneous Green's function between two locations in the subsurface using a single-sided boundary, without any artifacts from the focusing function, $\fxbxaw$.\\

\subsection{Implementation of Green's function retrieval}
\begin{figure}[!t]
\centering
\includegraphics[width=\columnwidth]{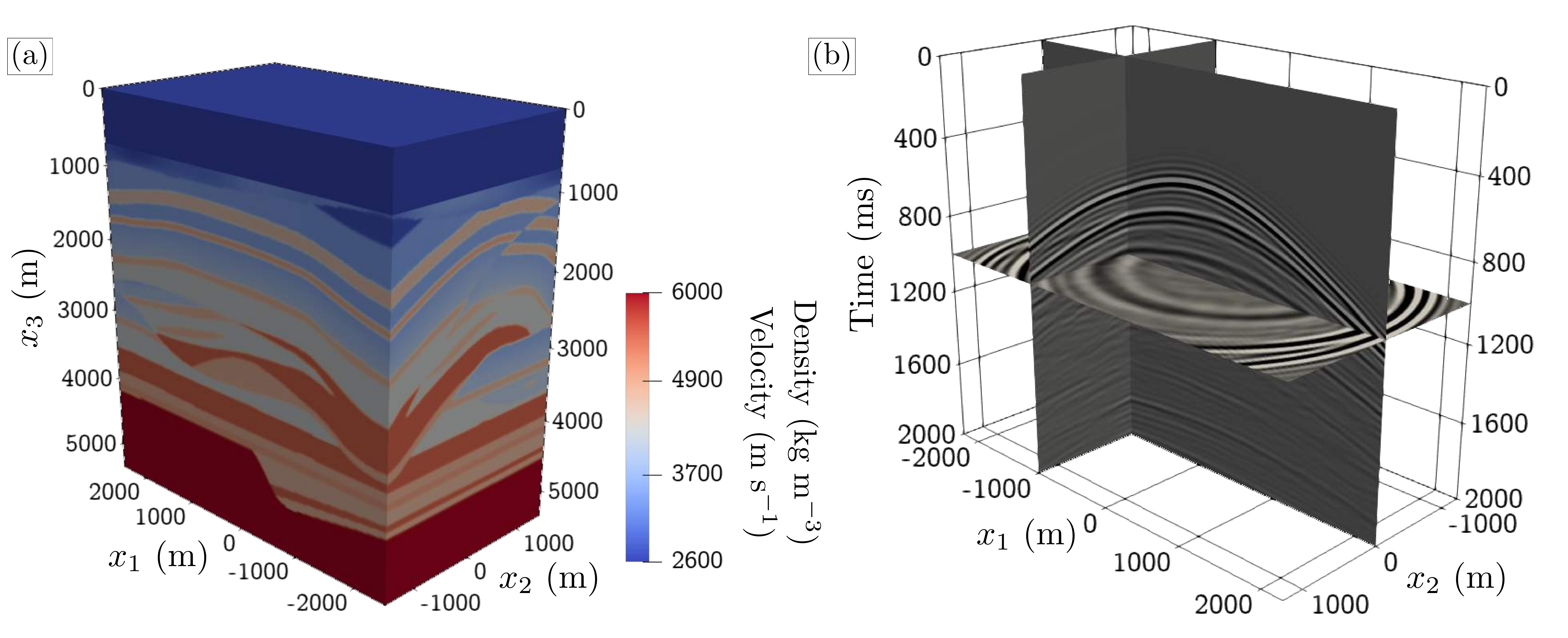}
\caption{(a) Velocity and density model of the subsurface, based on the Overthrust model by \cite{aminzadeh1997}. (b) Common-source record of a source, located at (0,0,0)m, recorded at the surface of the model in (a). The wavefield in (b) contains a wavelet that has a flat spectrum between 5 and 25 Hz.}
\label{recdata}
\end{figure}
We will demonstrate the results of the retrieval schemes in \eqnsref{gretclassapp}, (\ref{gret}) and (\ref{ghret}) with numerical examples. In order to obtain the required Green's functions and focusing functions on the right-hand sides of these equations, we employ the 3D Marchenko method on acoustic reflection data \cite{wapenaar2014marchenko}. The scheme allows one to retrieve the Green's function and focusing function between receivers at the surface of the Earth and a focal location in the subsurface of the Earth. To obtain these functions, a reflection response without surface related multiples at the surface of the Earth is required, as well as an estimation of the direct arrival from the surface of the Earth to the focal location. Usually, the time-reversed direct arrival of the Green's function from the focal location to the surface is used for this, even though this introduces errors proportional to the transmission losses into the final result \cite{thorbecke2017implementation}. 

\begin{figure}[!t]
\centering
\includegraphics[width=\columnwidth]{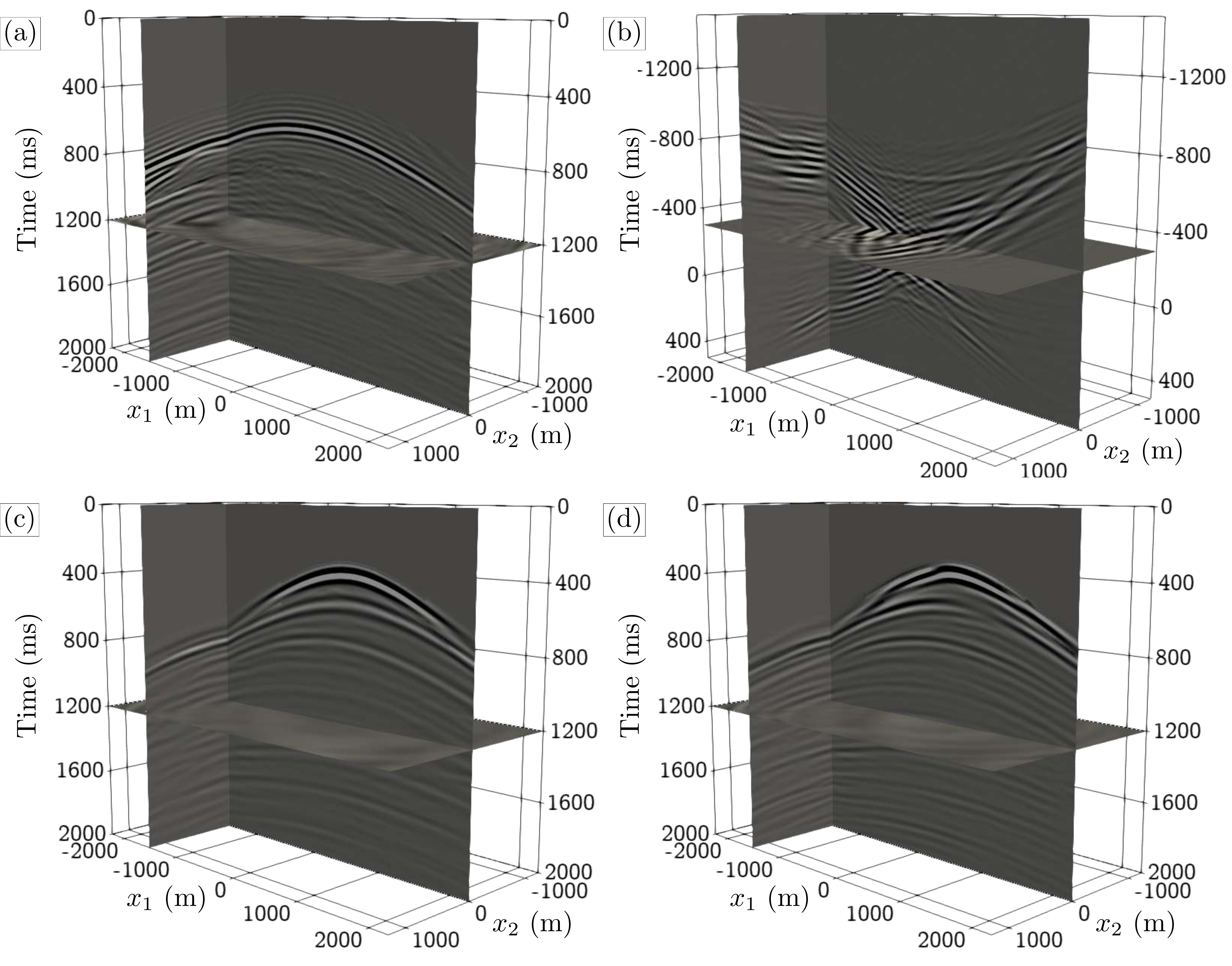}
\caption{Examples of wavefields obtained using the Marchenko method. (a) Green's function $\gxaxkw$ and (b) focusing function $\fpxkxaw-\{\fmxkxaw\}^*$ convolved with a wavelet with a flat spectrum between 5 and 25 Hz, with $\bxa=(-350,100,2150)$. Pressure wavefield $\pxkxbw$, i.e., a Green's function convolved with a 11Hz Ricker wavelet, with (c) an isotropic source and (d) pressure wavefield $\dopB{\pxkxbw}$ with a double-couple source, both with $\bxb=(500,-150,1025)$.}
\label{functionsch5}
\end{figure}

In this paper, we make use of our opensource 3D implementation of the Marchenko method \cite{thorbecke2019opensource}. To obtain the reflection data, we use a 3D finite-difference modeling code \cite{thorbecke2019opensource} together with a subset of the 3D Overthrust model \cite{aminzadeh1997}, which is shown in \figref{recdata}(a). To ensure strong reflections, the same model is used for the density and velocity values. To model the data, a fixed-spread acquisition is utilized, where a source is modeled at every receiver location. The source/receiver locations vary from -2250 to 2250m in the inline ($x_1$) direction, with a spacing of 25m, while the locations in the crossline ($x_2$) direction vary from -1250 to 1250m, with a spacing of 50m. We define a common-source record as the reflection response to a fixed source, observed by all receivers. The recording length of each common-source record is 4.0s with a temporal sampling of 4ms. An example of a common-source record is shown in \figref{recdata}(b). The data are modeled using a wavelet with a flat spectrum between 5 and 25 Hz. Examples of Green's functions and a focusing function obtained from these data and the Marchenko method can be found in \figref{functionsch5}.

Once we obtain the required Green's functions and focusing functions, we use them in the various retrieval schemes of \eqnsref{gretclassapp}, \eqref{gret} and \eqref{ghret}. To account for the fact that we don't have the analytical forms of the wavefields, we use numerical approximations of the schemes and make use of pressure wavefields with a band-limited source signature. We rewrite \eqnsref{gretclassapp}, (\ref{gret}) and (\ref{ghret}) as
\begin{equation} \label{gcanum}
\begin{split}
\phxaxbw&\\
\approx\sum^{n_R}_{j=1}\frac{2}{i\omega\rho_0}\Big\{ &\d3\gxaxkmw\pxkxbw\Big\}{\rm \Delta}^2\bx_j,
\end{split}
\end{equation}
\begin{equation} \label{gretnum}
\begin{split}
&\pxaxbw+\chi(\bxb)2is(\omega)\Im\{\fxbxaw\}\\
&=\sum^{n_R}_{j=1}\frac{2}{i\omega\rho_0}\pxkxbw\\
&\times\d3\Big(\fpxkxaw-\{\fmxkxaw\}^*\Big){\rm \Delta}^2\bx_j,
\end{split}
\end{equation}
\begin{equation} \label{ghretnum}
\begin{split}
&\phxaxbw=4\Re\sum^{n_R}_{j=1}\frac{1}{i\omega\rho_0}\pxkxbw\\
&\times\d3\Big(\fpxkxaw-\{\fmxkxaw\}^*\Big){\rm \Delta}^2\bx_j,
\end{split}
\end{equation}
where $\bxk$ is the location of the $j$th receiver at the surface of the Earth, $n_R$ is the amount of receivers and ${\rm \Delta}^2\bx_j$ indicates the receiver sampling distance. While ${\rm \Delta}^2\bx_j$ can be unique for each receiver position, in our fixed spread acquisition the value is the same for all receivers, namely ${\rm \Delta}^2\bx_j={\rm \Delta}x_1{\rm \Delta}x_2=25.0\cdot 50.0=1250$m$^2$. Note that in all the numerical representations, we have replaced $\ghxaxbw$, $\gxaxbw$ and $\gxxbw$ by $\phxaxbw$, $\pxaxbw$ and $\pxxbw$, respectively, while some of the other quantities are still denoted by their original symbol. In the application of \eqnsref{gcanum}-(\ref{ghretnum}), we assume that $\pxxbw$ is obtained either through the use of the Marchenko method or by a direct measurement, while $\gxaxw$, $\fpxxaw$ and $\fmxxaw$ are always obtained through the use of the Marchenko method. Therefore, we can control the source spectrum of the data that are used to generate the virtual receiver data. We ensure that $\gxaxw$, $\fpxxaw$ and $\fmxxaw$ have a source signature with a flat spectrum of amplitude 1.0 for a frequency range between 5 and 25Hz, so that the convolution with a unique source signature in that frequency range will produce the response to the latter source signal. The versions of $\phxaxbw$, $\pxaxbw$ and $\pxxbw$ that are used in \eqnsref{gcanum}-(\ref{ghretnum}) all include a 11Hz Ricker wavelet; an example of such a pressure wavefield can be found in \figref{functionsch5}(c), with its source at location $\bxb=(500,-150,1025)$. All other quantities are convolved with the wavelet with a flat spectrum between 5 and 25 Hz, similar to the reflection data from \figref{recdata}(b). Examples of a Green's function and focusing function convolved with such a wavelet are shown in \figref{functionsch5}(a) and (b), respectively, with their source present at location $\bxa=(-350,100,2150)$. The application of the band-limitation introduces one more complication, namely that \eqnref{ghretnum} is only valid if the source spectrum of $\pxaxbw$ is purely real valued, which holds for the source spectrum of the zero-phase Ricker wavelet. 

\begin{figure*}[!t]
\centering
\includegraphics[width=\columnwidth]{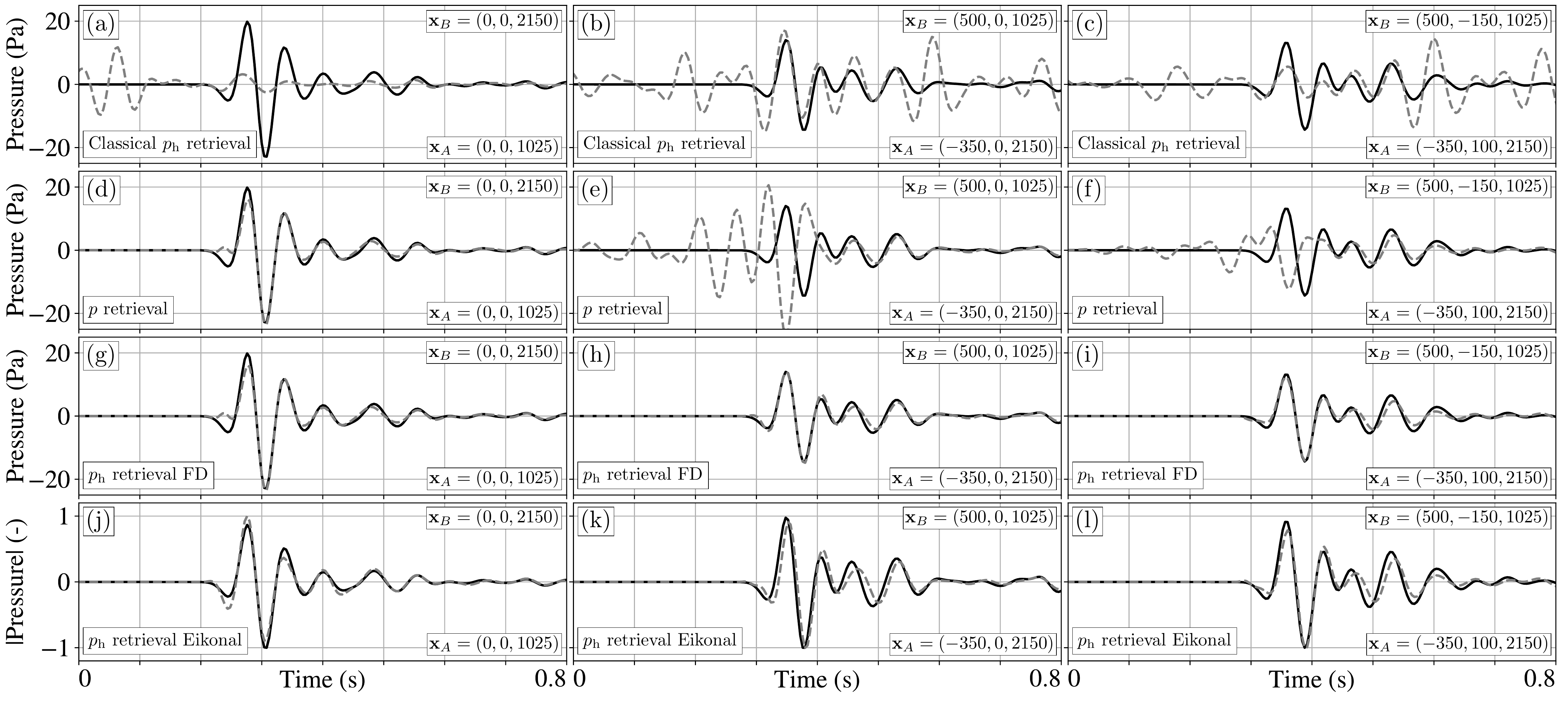}
\caption{Green's funcions of pairs of virtual sources and virtual receivers for different locations and different types of retrieval scheme. The solid black lines are the exact (directly modeled) Green's functions and the dashed gray lines are the retrieved functions. Each column corresponds to a different pair of locations. The first row corresponds to the classical retrieval scheme of \eqnref{gcanum}, the second row to the Marchenko retrieval scheme of \eqnref{gretnum} and the third row to the homogeneous Marchenko retrieval scheme of \eqnref{ghretnum}. For all of these rows, the first arrival required for the Marchenko method is obtained using finite-difference modeling. For the fourth row the same retrieval method is used as in the third row, except the first arrival is obtained using an Eikonal solver, instead of finite-difference modeling. The traces in the final row are all normalized. All traces contain a 11Hz Ricker wavelet.}
\label{traces}
\end{figure*}
To demonstrate the validity of our implementation, we show the result of the retrieval schemes in \figref{traces}, using the two-step method. Each column corresponds to a different pair of virtual source and virtual receiver positions, while each row corresponds to a different type of retrieval method. The first column has a virtual receiver located above the virtual source and the positions only differ in depth. In the second column the virtual receiver is located below the virtual source and the locations differ in both the inline direction and depth. For the third column the virtual receiver is located below the virtual source and the locations differ in all three principal directions. The required Green's function and focusing function are obtained using the Marchenko method and a first arrival that was obtained by modeling in the exact medium. We invert the first arrival instead of only time-reversing it, to compensate the transmission losses. While this is not a realistic scenario, as for field data we would not be able to use the exact model, we wish to demonstrate that the method is, at least in theory, capable of obtaining the exact amplitudes. The source has an isotropic radiation pattern. For each panel, the result that is obtained through the use of a retrieval scheme is plotted in dashed gray, while a directly modeled reference solution is shown in solid black.

The homogeneous wavefield that is obtained using \eqnref{gcanum} is shown in the first row of \figref{traces}. Both the Green's function for the virtual source and the virtual receiver were obtained using the Marchenko method. For all location pairs, the results are poor. While the order of magnitude of the retrieved wavefield is similar to that of the direct modeling, the amplitudes have a strong mismatch and there are artifacts present at all times. The exceptions are the first arrival in \figref{traces}(b) and the early coda in \figref{traces}(c). Aside from these events however, all other events are wrong and there are still significant artifacts for both examples. 

The second row shows the pressure wavefield that is obtained using the open boundary retrieval scheme from \eqnref{gretnum}. When the source is located below the virtual receiver, as is the case in \figref{traces}(d), the result shows a good match to the reference solution in both amplitude and arrival time. Because the first arrival is isolated, we apply a muting window to the data before the first arrival to remove numerical artifacts. This muting window is determined through the use of the smooth model, by using the Eikonal solver to determine the arrival time of the first event from the virtal source to the virtual receiver. A small shift is applied to account for the width of the wavelet. This arrival time is compared to the arrival time of the strongest amplitude in the retrieved causal Green's function, which should belong to the first arriving event of the causal Green's function. If these arrival times are similar, a single arrival time is determined and used as the cut-off time. A smooth taper is applied near the cut-off to avoid a sharp transition.

When the source is located above the receiver, the result degrades in quality. There are strong artifacts present in the result at times before the first arrival and the first arrival has the wrong polarity and amplitude. As these artifacts are of a similar magnitude as the first arrival, we cannot apply the muting window, because the arrival time of the strongest amplitude no longer matches the arrival time estimated by the Eikonal solver. The retrieved coda in these two latter cases is still accurate, with some slight mismatch in the amplitude of the events. This is caused by the different lateral positions of the source and receiver. The aperture and sampling of the data in both the inline and crossline direction are limited, so the exact events become harder to obtain. The overall coda shows a good match to the reference. 

To improve the results of the retrieval, the representation from \eqnref{ghretnum} is used to retrieve the homogeneous Green's function, shown in the third row of \figref{traces}. The results in \figref{traces}(d) and (g) are identical, which corresponds to the condition in \eqnref{chixb}. The improvement is apparent when the source is located above the receiver as is the case in \figref{traces}(h) and (i). Compared to the results in \figref{traces}(e) and (f), the unwanted artifacts are removed and the first arrival is retrieved properly. Here, we once again apply the muting window before the first arrival, because there is a match between the estimated first arrival time of the Eikonal solver and the strongest amplitude. The amplitude mismatch in the coda is still present, indicating that this is a limitation caused by the aperture of the recording array and not of the type of retrieval method.

Finally, in the bottom row of \figref{traces}, we apply \eqnref{ghretnum} again, however, this time the first arrivals used in the Marchenko method are obtained using an Eikonal solver in a smoothed version of the velocity model. The Eikonal solver calculates the arrival time of the first event in the wavefield \cite{Vidale1990}, as well as a geometrical spreading factor to estimate the amplitude along the wavefront \cite{spetzler2005ray}. The choice to utilize the Eikonal solver is made  to simulate a more realistic situation, where accurate model information would not be available and the estimate of the first arrival using finite-difference modeling would also be not exact. Because the exact amplitude of the first arrival cannot be obtained when a smooth velocity model is used, the retrieved homogeneous Green's function is normalized and compared to a normalized version of the reference solution. This is intended to show that even when the exact amplitude cannot be obtained, the relative amplitude can be properly obtained. The matches for all three source-receiver pairs are good, but of a lesser quality than when the finite-difference modeling is employed. Due to the complexity of the model, as well as the smoothing, the Eikonal solver can encounter issues with obtaining the correct arrival times. Furthermore, we only use an estimation of the amplitude distribution along the wavefront, which also will not properly represent the true effect that the subsurface would have on the amplitude. However, the results still support that use of an Eikonal solver for 3D media can yield useful results.

\subsection{Visualization of the 3D results}
\begin{figure*}[p]
\centering
\includegraphics[width=0.75\columnwidth]{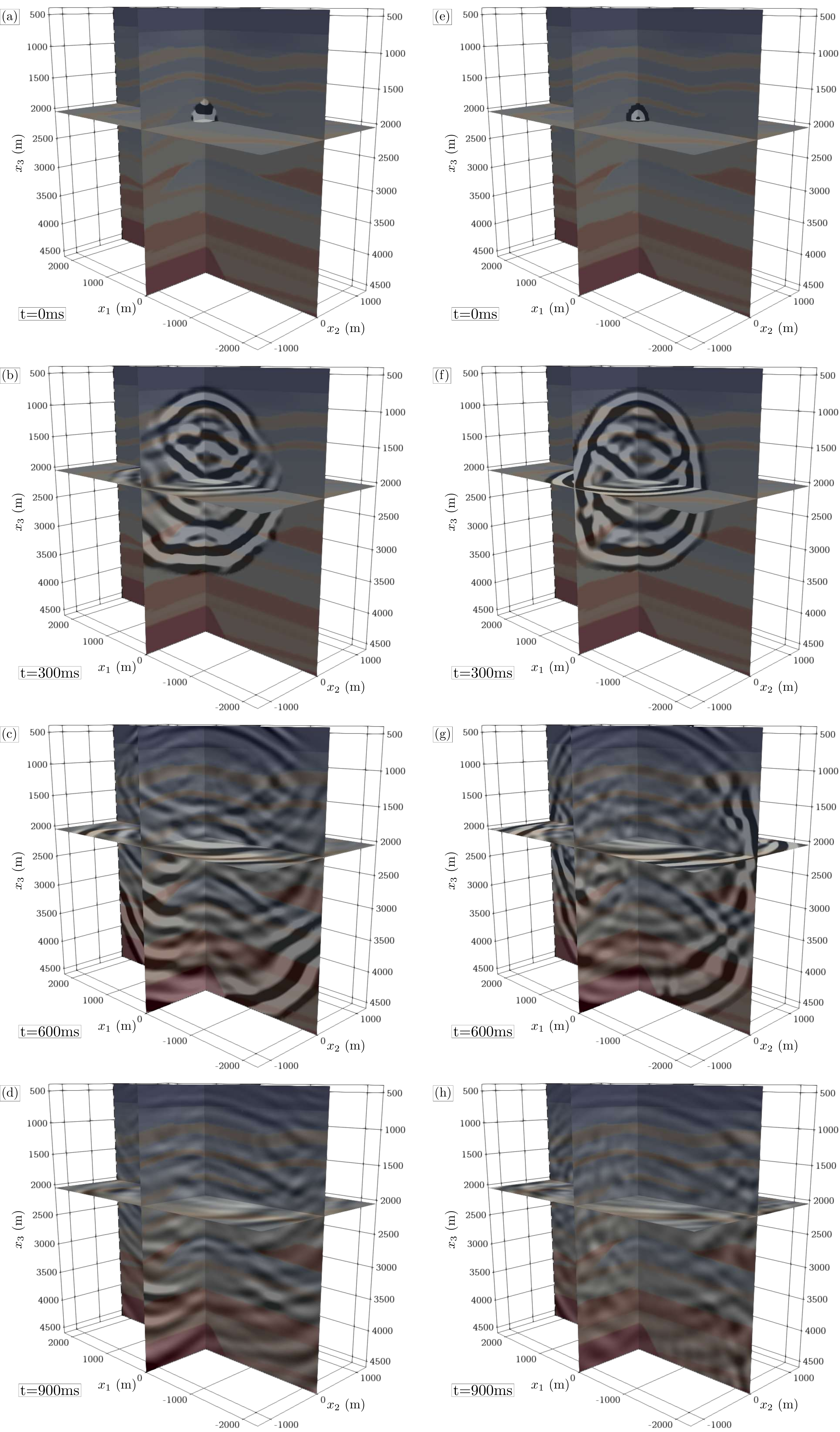}
\caption{3D snapshots of the homogeneous Green's function retrieval of an isotropic virtual source in the Overthrust model at (a) 0ms, (b) 300ms, (c) 600 ms and (d) 900ms. For comparison, (e), (f), (g) and (h) are snapshots of a directly modeled homogeneous pressure wavefield at 0ms, 300ms, 600ms and 900ms, respectively. The source is located at $\bxb=(0,0,2050)$m. The first arrivals for every virtual source-receiver pair were obtained using the Eikonal solver, similar to the results in \figsref{traces}(j)-(l). All wavefields contain a 11Hz Ricker wavelet and contain an overlay of a cross section of the Overthrust model to indicate the locations where we expect scattering to take place. Both columns are independently clipped at a tenth of the maximum value of the total wavefield. The Pearson Correlation Coefficient between the two wavefields is 0.542.}
\label{monopole}
\end{figure*}
While the traces in \figref{traces} demonstrate the validity of our approach, they are limited in scope. To further test our approach in 3D, we obtain the results for not just a single source-receiver pair. Instead we retrieve a large amount of focusing functions and use these in \eqnref{ghretnum} to visualize the retrieved Green's functions evolving in time through the 3D medium. To obtain the results, we use the approach employing the Eikonal solver, similarly to how we obtained the results in the bottom row of \figref{traces}. In this case, we do not apply normalization, as we do not directly compare the result of the Eikonal solver to a reference result. The retrieved amplitudes are not exact, however, the relative amplitude differences between different locations will be similar to those of the true wavefield. This is because the amplitude estimation is calculated independently for each virtual receiver location. The reason we use an Eikonal solver to model the first arrivals is that the computational load and storage space for the use of finite-difference modeling are not feasible for the amount of source-receiver pairs that we desire. The use of the Eikonal solver is a similar approach as was used in previous research \cite{brackenhoff2019}, however, in this work, we extend its use to 3D. We use the Eikonal solver to obtain the first arriving event of the focusing functions at locations along three slices through the 3D medium, one with a fixed depth at 2050m with an inline and crossline position from -2250 to 2250m and -1250 to 1250m respectively, one at a fixed inline position of 0m with a depth and crossline position from 400 to 4600m and -1250 to 1250m respectively and one slice at a fixed crossline position of 0m with a depth and inline position from 400 to 4600m and -2250 to 2250m, respectively. For all slices, the sampling in the depth, inline and crossline direction is 25, 25 and 50m, respectively. For the source wavefield $\pxkxbt$, we obtain a single pressure wavefield due to a source with an isotropic radiation pattern at $\bxb=(0,0,2050)$m.

The results of the retrieval using these data and \eqnref{ghretnum} are shown in the left column of \figref{monopole}. For comparison, we have created a reference homogeneous pressure wavefield by modeling the wavefield directly in the exact medium and superposing it with its time-reversal. The two wavefields are clipped at different values, because of the amplitude difference. Both clipping factors were set to a tenth of the maximum value of their respective wavefields. We also apply the muting window before the first arrival at all the positions. This is once again done by comparing the estimated arrival times of the first arriving event from the virtual source to the virtual receiver through the use of the Eikonal solver and finding the strongest amplitude. To avoid outliers, we also compare the arrival times of nearby virtual receivers. Near the edges, we observe some of these outliers. The four rows each correspond to a different moment in time, namely 0, 300, 600 and 900ms. When comparing the results of the retrieval to the direct modeling, it can be seen that while the match in certain locations is strong, in other locations events appear to be missing. This is due to the finite aperture of the data. The theoretical representations in \eqnsref{gret} and \eqref{ghret} assume that the aperture of the data is infinite. In reality, the aperture is limited, especially in the crossline direction. The events in the homogeneous Green's function are reconstructed from the reflection data, so if an event is not present in the reflection data, it will not be reconstructed properly. The horizontally traveling wavefield, especially near the edges of the aperture, will not arrive at the surface within the range of the aperture. The deeper the target is in the medium, the more severe this problem can become. This effect is particularly apparent in \figref{monopole}(a), where the difference with the direct modeling is very strong. The wavefields that travel away from the source along the horizontal are not reconstructed, hence the apparent difference in the source radiation pattern. It should be noted that if the velocity of the medium is increasing with depth, the refraction of the waves will ensure that more angles of the wavefield arrive at the surface of the medium. In the Overthrust model, the propagation velocity of the medium is generally increasing with depth, however, there are some low velocity zones present at greater depths. Because of the general increasing trend, some of the horizontally traveling wavefield at greater depths is still recovered. 

The part of the wavefield that is traveling at a smaller angle is reconstructed properly, even at large depths and at the edges of the aperture. The events in the center of the model are reconstructed properly. The amplitudes and arrival times of the events are not correct everywhere, which is caused by the use of a smooth velocity model and the Eikonal solver for the direct arrivals, instead of modeling these in the exact medium. To give a more quantitative result for the accuracy of the retrieval, we employ the Pearson Correlation Coefficient (PCC) \cite{benesty2009pearson}. This coefficient ranges from -1 to 1. A value close to $\pm 1$ indicates strong correlation, while a value close to 0 indicates weak or no correlation. The polarity of the coefficient indicates whether the correlation is positive or negative. The PCC between the columns of \figref{monopole} is 0.542, which indicates medium correlation. The relatively low correlation value is likely caused by the issues discussed previously. However, the results and the PCC still show the potential of the Marchenko method for 3D virtual seismology.

\section{Moment tensor monitoring}
\subsection{Non-isotropic point source}
In reality, an event in the subsurface is seldom generated by an isotropic point source. Instead the source wavefield is often caused by faulting, the mechanism of which can be described by a moment tensor \cite{aki2002quantitative}, which causes the amplitude along the wavefront to vary. The double-couple source mechanism is often used, which is a moment tensor that describes a pure shear fault, by its strike, rake and dip \cite{li2014global}. In previous work, the double-couple source mechanism was combined with the Marchenko method to obtain the virtual response of a double-couple point source in the subsurface, as well as that of a rupture plane \cite{brackenhoff2019virtual}. Here, we wish to demonstrate that similar results can be achieved in 3D. We repeat the examples of the isotropic point source, using \eqnsref{gcanum}-\eqref{ghretnum}, however, we replace the isotropic source at $\bxb$ in $\pxkxbw$ by a double-couple source generated by a moment tensor. We use an operator $\dopB{\cdot}$, which transforms the radiation pattern of the source at $\bxb$ from an isotropic radiation pattern to a double-couple radiation pattern. It is defined as
\begin{equation} \label{dcoup}
\dopB{\cdot}=(\theta_i^\parallel+\theta_i^\perp)\partial_{i,B},
\end{equation}
where $\partial_{i,B}$ is a component of the vector containing the partial derivatives acting on the monopole signal originating from source location $\bxb$, which alters the radiation pattern, $\theta_i^\parallel$ is a component of a vector that orients one couple of the signal parallel to the fault plane and $\theta_i^\perp$ is a component of a vector that orients the other couple perpendicular to the fault plane. Because we are dealing with acoustic reflection data, we only model the P-waves of the double-couple source, select the first arrival and use it in the Marchenko method to obtain the desired virtual double-couple response $\dopB{\pxkxbw}$. An example of such a wavefield can be found in \figref{functionsch5}(d), which has a source at the same position as the pressure wavefield with an isotropic source in (c). This wavefield is then used in \eqnref{gcanum} or \eqref{ghretnum} to obtain $\dopB{\phxaxbw}$ or in \eqnref{gretnum} to obtain $\dopB{\pxaxbw}$. Previous research has suggested that the double-couple source is not always a sufficient description of an earthquake source \cite{julian1998non}. Our wavefield retrieval method is valid for any type of moment tensor, however, for the sake of simplicity, we stick with the double-couple representation.

\begin{figure*}
\centering
\includegraphics[width=\columnwidth]{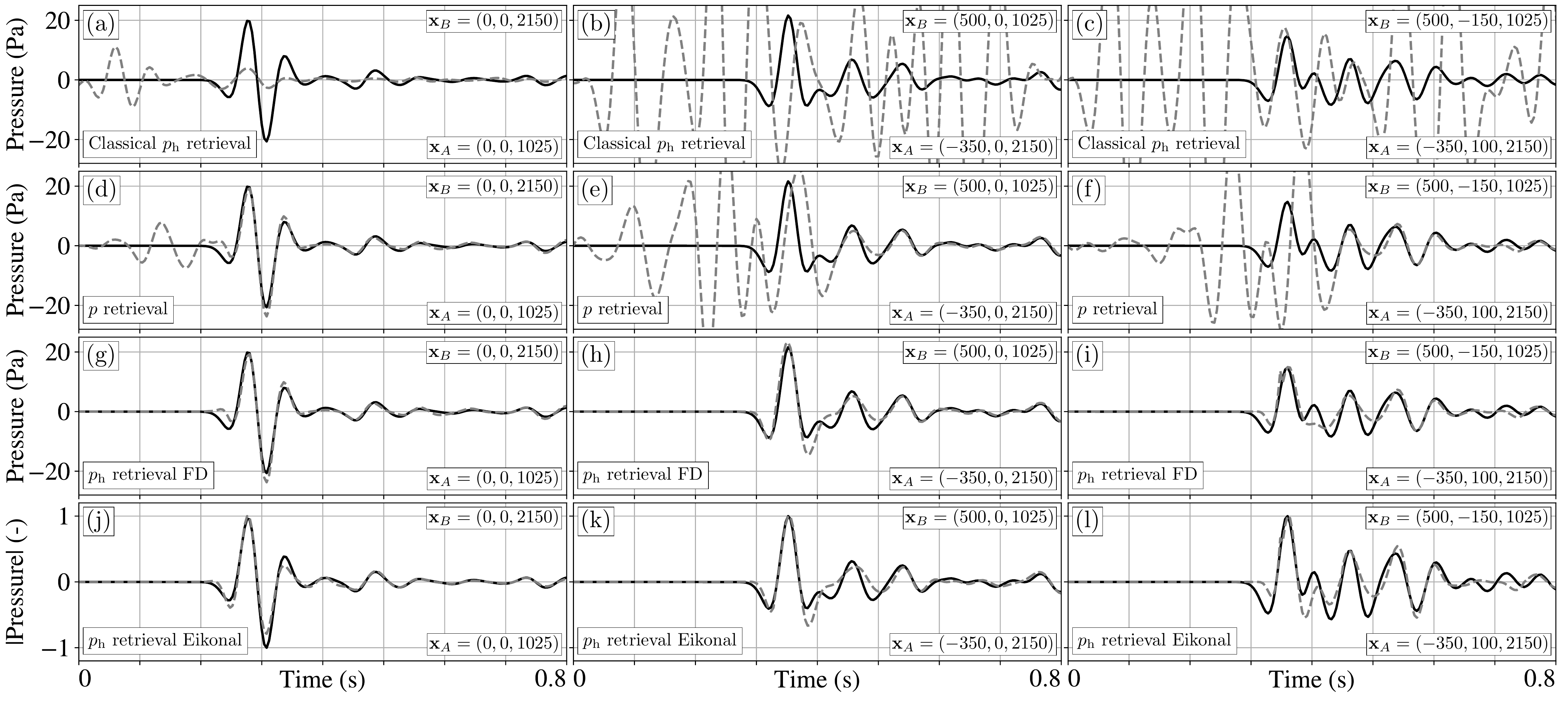}
\caption{As \figref{traces}, but for a virtual double-couple source with  a strike, rake and dip of 19, 68 and 25 degrees, respectively.}
\label{tracesdc}
\end{figure*}

For our example, we use a double-couple source with a strike, rake and dip of, 19, 68 and 25 degrees, respectively, and obtain the response between the virtual source-receiver pairs, similar to the examples in \figref{traces}, using the same color scheme as in that figure. The results are shown in \figref{tracesdc}, using the same setup as \figref{traces}. The first, second and third row show the retrieval using \eqnref{gcanum}, \eqnref{gretnum} and \eqnref{ghretnum} using finite-difference modeling in the exact medium, respectively, while the fourth row shows the retrieval using \eqnref{ghretnum} and the Eikonal solver. The columns all show different source-receiver pairs, at the same locations as were used for \figref{traces}. 

The results in \figref{tracesdc} lead us to similar conclusions as the results in \figref{traces}, the only major difference is in the shape and amplitude of the events, caused by the different source mechanisms. The classical representation contains major artifacts and the single-sided retrieval of the Green's function contains artifacts if the virtual receiver is located below the virtual source. If the single-sided representation for the homogeneous Green's function is employed, the result is accurate in all cases. The use of the Eikonal solver instead of finite-difference modeling affects the absolute amplitude but not the relative amplitudes. There is a limit in the accuracy of the results, caused by the finite aperture of the acquisition.

\begin{figure*}[p]
\centering
\includegraphics[width=0.75\columnwidth]{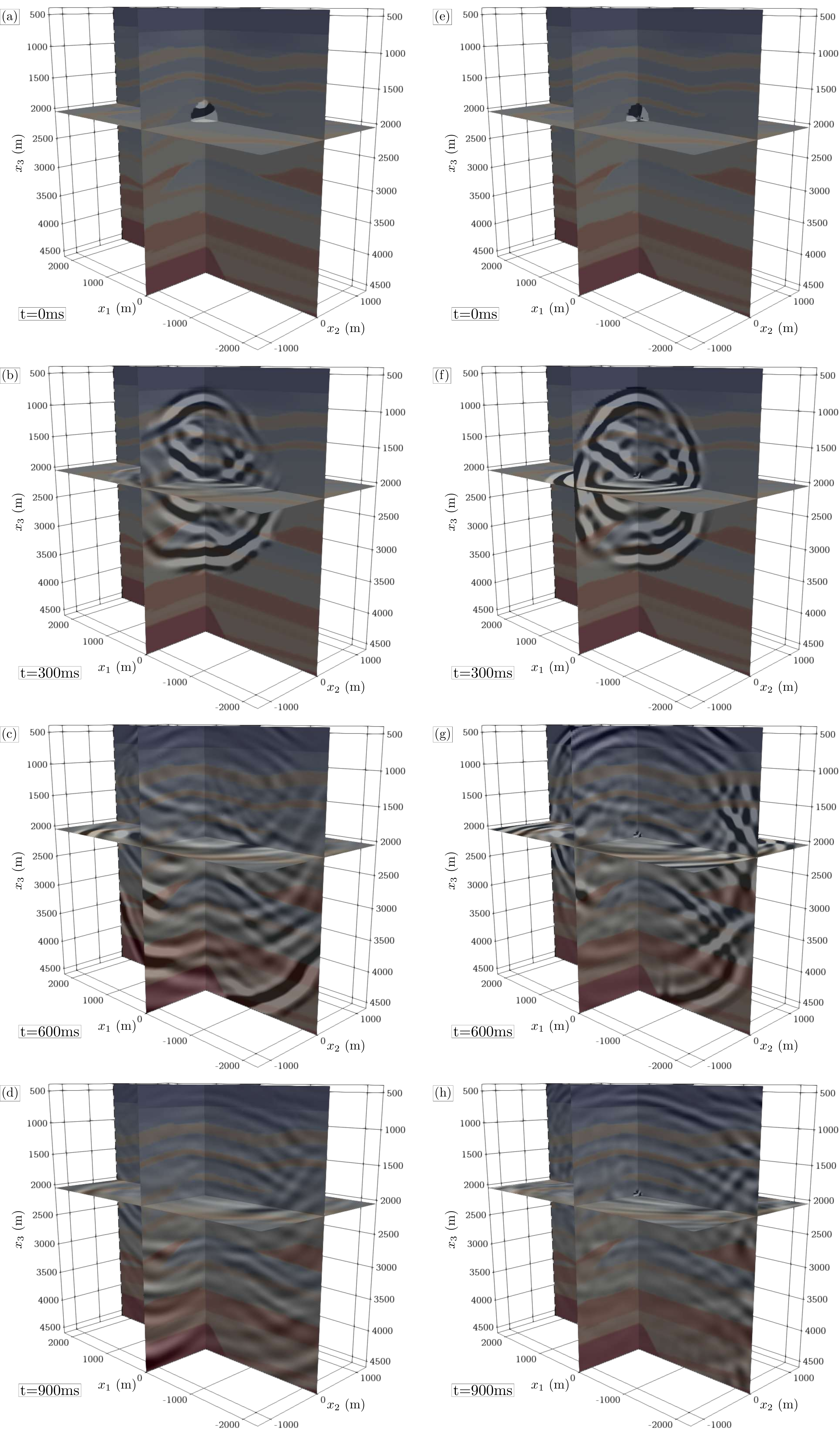}
\caption{As \figref{monopole}, but for a virtual double-couple source with  a strike, rake and dip of 19, 68 and 25 degrees, respectively. Note that the direct modeling contains a source artifacts caused by only modeling the P-waves. The Pearson Correlation Coefficient between the two wavefields is 0.433. The PCC increases to 0.504 when the source region is excluded.}
\label{doublecouple}
\end{figure*}

To further investigate the effects of using a double-couple source mechanism, we retrieve the wavefield for the same three 3D slices as we did for \figref{monopole}. The result is shown in \figref{doublecouple}, where the left column shows the result for the retrieval using the Eikonal solver and \eqnref{ghretnum}, while a direct modeling is shown in the right column for comparison. The clipping values are once again set to a tenth of the maximum value of the respective wavefields. Note that the direct modeling contains a source artifact caused by the modeling of only the P-waves. When the results of the retrieval and the direct modeling are compared, most of the nearly vertically traveling events are properly retrieved, not only in arrival time, but also in polarity. For events traveling nearly horizontally, the retrieval is once again poorer. The PCC is also retrieved between the columns in this figure, which results in a coefficient of 0.433. The lower value is caused by the source artifact as the PCC rises to 0.504 when this region is excluded. The correlation indicates that the quality of the retrieval is similar to that of the monopole source. Overall, the results using the double-couple source have a similar quality as the results for the isotropic source, which demonstrates that in 3D, the double-couple source can be successfully integrated into the homogeneous Green's function retrieval method.
\subsection{Rupture}
In previous sections, we have only considered point sources, however, in the field, an earthquake is seldom a single event, rather, it consists of a cluster of several events that are activated over an area for a period of time \cite{schorlemmer2005variations}. Hence, the total wavefield of an earthquake is not the result of a single instantaneous source, instead, it consists of a superposition of wavefields caused by different sources that are activated at different times. To approximate this kind of wavefield, we define a total wavefield $\Pxat$ that consists of a superposition of wavefields that are caused by double-couple point sources. The superposition can be expressed as
\begin{equation} \label{Ptott}
\begin{split}
&\Pxat=\sum^{n_S}_{k=1} \dopBk{\pxaxbjt}\\
&=\sum^{n_S}_{k=1} \int_{-\infty}^{\infty}\dopBk{G(\bxa,\bxbj,t-t')} s^{(k)}(t'){\rm d}t',
\end{split}
\end{equation}
where $\bxbj$ indicates the location of the $k$th source of a total of $n_S$ sources, $\dopBk{\cdot}$ is the double-couple operator for each location and $s^{(k)}(t)$ is the corresponding source signal for each location that contains all the information for the source strength, activation time and duration. Because of the different activation times, the source spectrum of $\Pxat$ is no longer purely real-valued and can therefore not be used in \eqnref{ghretnum}. However, using it in \eqnref{gretnum} is still valid, as no time-reversal is applied. We rewrite \eqnref{gretnum} for this purpose as
\begin{equation} \label{Pret}
\begin{split}
&\Pxaw\\
&+\sum^{n_S}_{k=1}\dopBk{\chi(\bxbj)2is^{(k)}(\omega)\Im\{\fxbjxaw\}}\\
&=\sum^{n_R}_{j=1}\frac{2}{i\omega\rho_0}\sum^{n_S}_{k=1}\dopBk{\pxkxbjw}\\
&\times\d3\Big(\fpxkxaw-\{\fmxkxaw\}^*\Big)\Delta^2\bx_j.
\end{split}
\end{equation}
In \eqnref{Pret}, we can retrieve $\Pxaw$, however, we will also obtain the focusing function artifacts that are related to each source position, below the source depth. As there are multiple sources, that can have different depths, the artifacts related to one source can interfere with the part of the signal that originates from a deeper part of the medium. Consequently, only above the shallowest source depth can we expect to obtain the correct wavefield at all times. For deeper parts of the medium, we expect to retrieve artifacts before and around the first arrival time and the correct coda at later times, similar to the results that were shown in \figsref{traces} and \ref{tracesdc}. Obtaining the wavefield in this way is a one-step process, where we first measure the total wavefield of an actual rupture and use it in combination with the focusing functions, obtained through the Marchenko method, to monitor the subsurface with virtual receivers. Because of this, we refer to this method as a monitoring approach. Note that for this approach, one requires not only the reflection data at the surface of the Earth, but also the measurement of a wavefield that was caused by a rupture in the subsurface. 

On the other hand, to obtain the response to a virtual rupture, we can use a two-step process to retrieve $\Pxat$. We call it a two-step process because the Marchenko method is applied twice. Instead of measuring the resulting wavefield of the superposed sources, we use the Marchenko method to retrieve the individual wavefields $\dopBk{\phxaxbjw}$ related to each source position. In this case we do not measure the total wavefield, but predict it by using the Marchenko method to obtain the source wavefield $\dopBk{\pxkxbjw}$ before using it in \eqnref{ghretnum}. Because of this, we can ensure that the source spectrum of $\dopBk{\pxkxbjw}$ for each individual virtual source is purely real-valued, before we apply the time-reversal. The wavefields that we retrieve in this way are free of the artifacts related to the focusing function and can be combined to form $\Pxat$:
\begin{equation} \label{Phret}
\begin{split}
&\Pxat\\
&=\sum^{n_S}_{k=1} H(t-t^{(k)})\dopBk{\phxaxbjtt},
\end{split}
\end{equation}
where $H$ is the Heaviside function and $t^{(k)}$ is the activation time of the source. In \eqnref{Phret}, we shift the signals in time by $t^{(k)}$ before superposition is applied. Because these wavefields are time-shifted and homogeneous, i.e. they contain time-shifted versions of $\dopBk{\pxaxbjt}$ and $\dopBk{\pxaxbjmt}$, the acausal part of one wavefield may interfere with the causal part of another wavefield. The Heaviside function is applied to remove all acausal parts of the wavefields to avoid such an issue. While this approach cannot be used for the monitoring of wavefields measured in the field that are caused by sources that are active over a period of time, the approach can be used to forecast the total wavefield of a virtual rupture, given a specific distribution of sources. Hence, we refer to this approach as the forecasting approach. Unlike the monitoring approach, this approach only requires reflection data at the surface of the Earth and nothing else. The forecasting approach can be used to predict the propagation of the wavefield caused by a possible rupture in the subsurface.

\begin{figure}
\centering
\includegraphics[width=0.7\columnwidth]{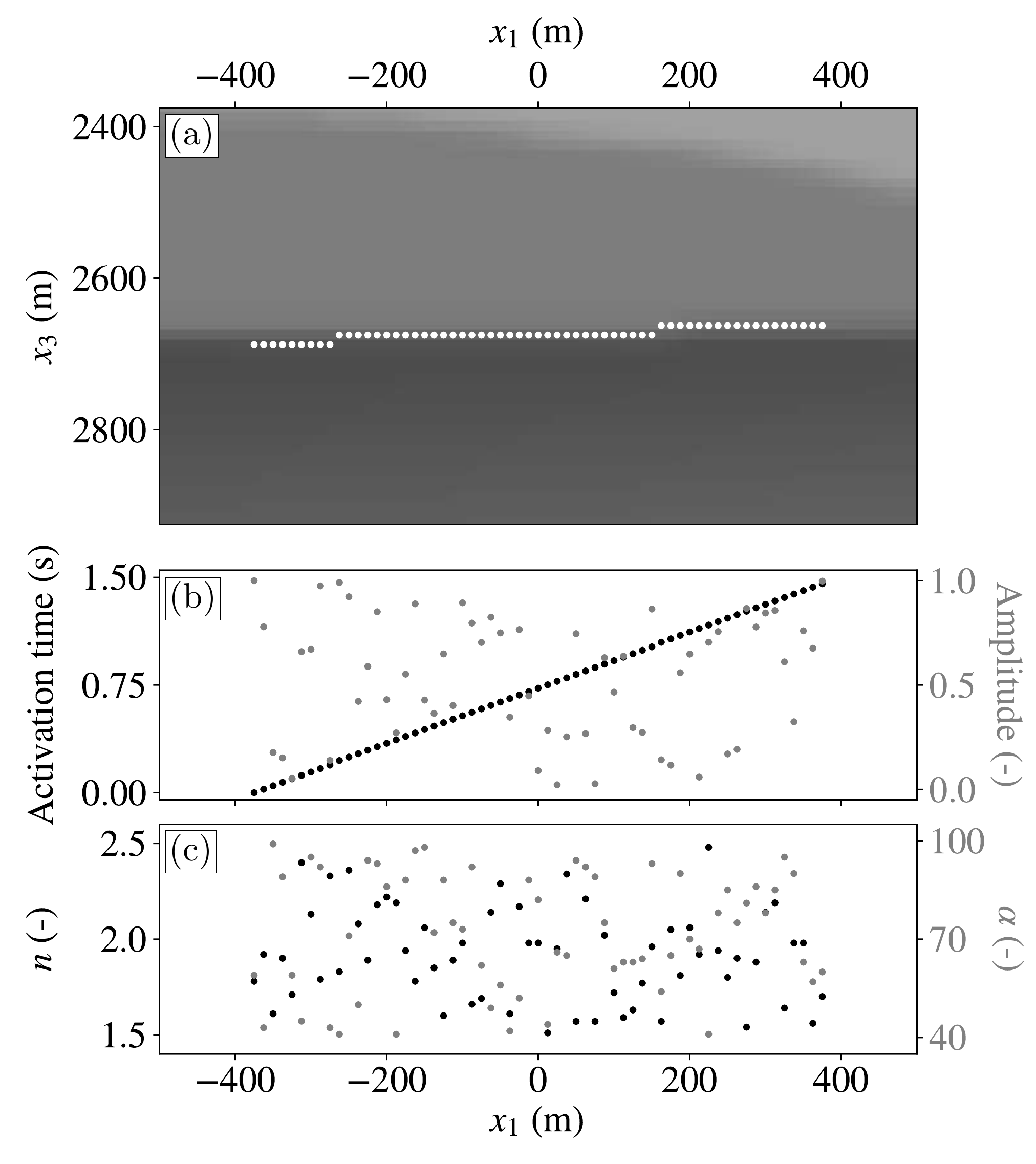}
\caption{(a) Locations of individual double-couple sources with a strike and rake of 90 and 0 degrees, respectively. The dip is oriented along the fault direction for each location. The slice is located along a constant crossline position of 0m. (b) Activation time and amplitude in black and gray, respectively, and (c) time exponent $n$ and exponential decay factor $\alpha$ in black and gray, respectively, for computing the Berlage wavelets using \eqnref{berlage}. The horizontal positions of the sources in (a), (b) and (c) match.}
\label{dcinfo}
\end{figure}

To demonstrate the monitoring and forecasting of the total wavefield, we consider a rupture plane in the Overthrust model that consists of a cluster of 61 point sources with a double-couple radiation pattern and that are activated at different points in time. Instead of retrieving wavefields that contain the zero-phase wavelet, like we have done in the previous examples, we retrieve wavefields that contain a unique causal wavelet for each source position, as wavefields in the real subsurface will be causal and not zero-phase. We choose the Berlage wavelet, which is defined as \cite{aldridge1990berlage}:
\begin{equation} \label{berlage}
W(t)=AH(t)t^ne^{-\alpha t}\cos(2\pi f_0t+\phi_0),
\end{equation}
where $A$ is the amplitude of the wavelet. The time exponent $n$, exponential decay factor $\alpha$, initial phase angle $\phi_0$ and peak frequency $f_0$ control the shape of the wavelet. To ensure that the wavelet has an amplitude equal to zero at $t=0$, we use an initial phase angle of -90 degrees. For the peak frequency, we use the same peak frequency as we used for the Ricker wavelet, namely 11Hz. However, for the amplitude, time exponent and exponential decay factor, we take random values, to simulate a heterogeneous region along the rupture plane. The schematic overview for the rupture simulation can be found in \figref{dcinfo}. The sources are located along a fault in the model, where each source has a strike and rake of 90 and 0 degrees, respectively, and is located at a fixed crossline position of 0m. The dip of the source is dictated by the fault orientation at each source location. \figref{dcinfo}(a) contains the locations of the sources, while \figref{dcinfo}(b) shows the activation time and random amplitude and \figref{dcinfo}(c) shows the random time exponent and exponential decay factor that are used for the Berlage wavelets. The activation time for the sources is linear, with a time delay of 24ms between the activation of subsequent sources, except for the positions where the depth of the source changes. In these cases the time delay is increased to 32ms to account for the increase in step size. In this way, we simulate a rupture activating and propagating along the rupture plane with a velocity of 520m s$^{-1}$.

\begin{figure*}[p]
\centering
\includegraphics[width=0.75\columnwidth]{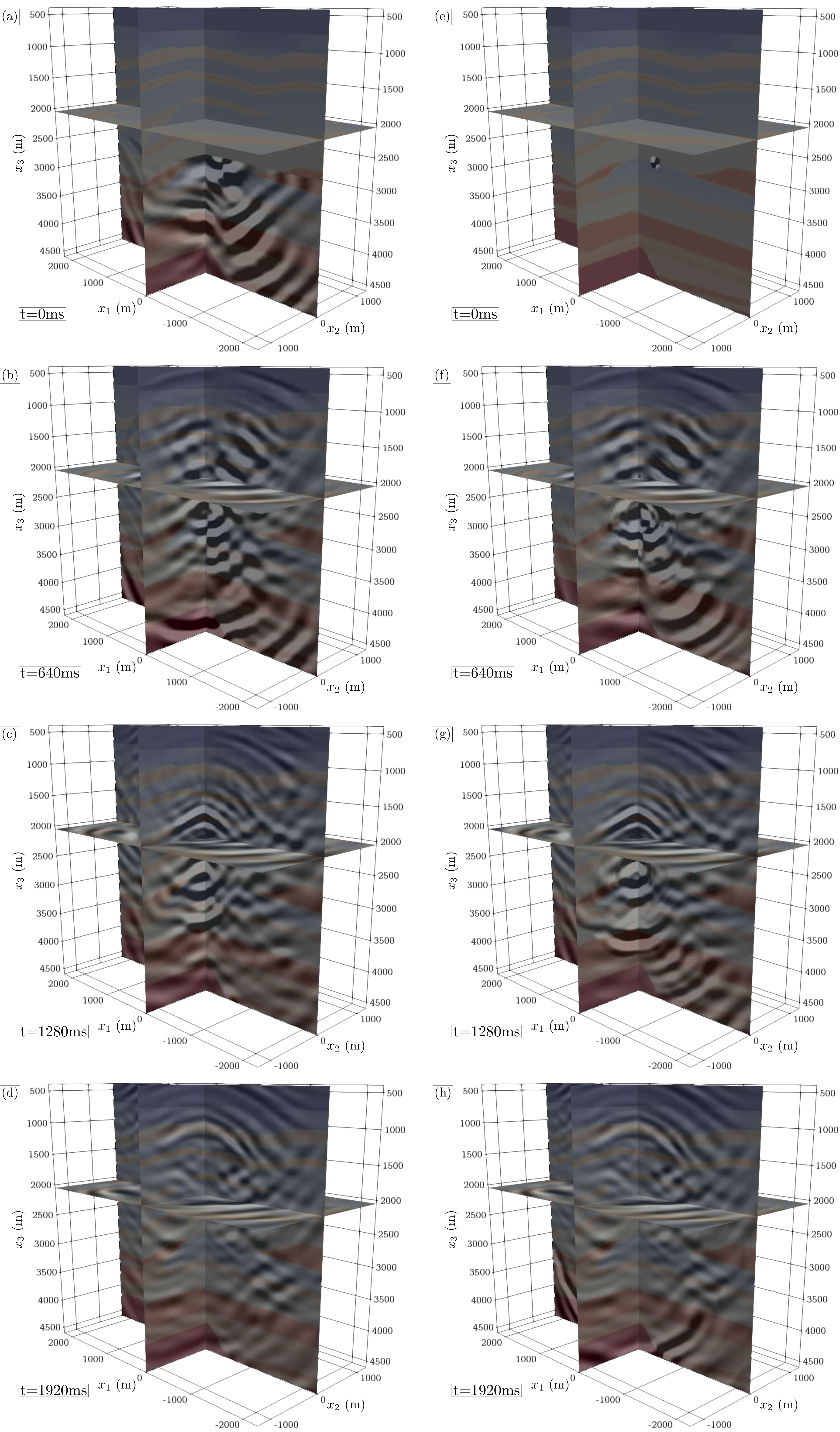}
\caption{3D snapshots of the Green's function retrieval for a wavefield caused by a rupture in the Overthrust model at (a) 0ms, (b) 640ms, (c) 1280 ms and (d) 1920 ms, using \eqnref{Pret}, and 3D snapshots of superposed and time-shifted wavefields in the Overthrust model, obtained using homogeneous Green's function retrieval using \eqnref{Phret}, at (e) 0ms, (f) 640ms, (g) 1280 ms and (h) 1920 ms. All wavefields have an overlay of a cross section of the Overthrust model to indicate the locations where we expect scattering to take place. Details about the locations, activation times and the wavelets of each source can be found in \figref{dcinfo}. The columns are clipped at the same value. The Pearson Correlation Coefficient between the two wavefields is equal to 0.389 for the entire duration, for the part of the wavefield after 640ms the PCC is 0.512 and after 1280ms it becomes 0.836.}
\label{rupture}
\end{figure*}

The results of both the one-step and the two-step process can be found in \figref{rupture}. The left column of this figure shows the result for the one-step process for monitoring a signal, using \eqnref{Pret}, and the right column shows the result for the two-step process for forecasting a signal, using \eqnref{Phret}. Both columns are clipped at the same value. For the monitoring process, we convolve the Berlage wavelets with the source wavefield before we employ the causal Green's function retrieval. For the forecasting process, we use a flat spectrum wavelet to obtain the individual homogeneous Green's functions. Because we are creating everything from the data, this is a valid approach. After we obtain these homogeneous Green's functions, we convolve the functions with the Berlage wavelets, similar to \eqnref{phdeft}, to obtain the homogeneous wavefields. These wavefields are then utilized in \eqnref{Phret}. In both cases we apply the muting window again, in a similar way as was done for \figsref{monopole} and \ref{doublecouple}. When we apply this window to the forecasting approach, we mute the wavefields $\dopBk{\phxaxbjtt}$ before they are superposed using \eqnref{Phret}. The result for the monitoring approach is only muted at the depths above the shallowest source, because below these depths, the match between the arrival times of the first event estimated by the Eikonal solver and the strongest amplitudes is poor. The differences between the estimated first arrival times of adjacent virtual receivers is significant at these depths, so no general trend can be determined.

When comparing the results, it can be seen that, at 640ms, there is a strong difference between the monitoring and forecasting of the signal. Below the depth of the shallowest source location, the wavefield contains strong artifacts, which is consistent with the theory, however, above this depth, the wavefields of the two approaches are exactly the same. For later times, around 1280ms, the area below the shallowest source matches more between the two approaches, however, the deeper parts of the medium still shows significant differences. At 1920ms, the match between the two results is even closer, only the deepest parts of the model still contains artifacts for the monitoring approach. We showed that applying homogeneous Green's function retrieval for a single source is accurate, so the superposition of the homogeneous pressure wavefields yields a good result. While the monitoring approach does contain artifacts, we can use the method to monitor the wavefield in the subsurface between the surface and the shallowest source depth accurately. Moreover, we can also use this approach to obtain the coda of the signal for late times at all depths. We also obtain the PCC between the results of the monitoring approach and the forecasting approach, to obtain a measure for the similarity of the results. The PCC for the entire signal is equal to 0.389, caused by the strong differences in the results at early times. When the data before 640ms are removed, the PCC increases to 0.512, and when the data before 1280ms are removed, the PCC increases to 0.836, showing that the codas of the two wavefields have very strong correlation. Overall, the results support the potential of using the single-sided Green's function retrieval in 3D in the field.

\section{Discussion}
The results of the application of the single-sided representation to numerical data show a potential for the same application on field data. To utilize the single-sided representation, the Marchenko method has to be employed, which requires high quality reflection data. First of all, the data should not contain free-surface multiples, which is not an unrealistic demand, as there exist many ways to remove the free-surface multiples from reflection data \cite{verschuur1992adaptive,van2009estimating}. Additionally, the data should be well-sampled in both the inline and the crossline direction. While the inline direction is usually densely sampled, the crossline direction is often more sparsely sampled. Previous publications have shown that the crossline spacing can be made more dense through the use of interpolation and that the Marchenko method can be employed using these interpolated data \cite{staring2020narrow}. 

When the reflection response is densely sampled, applying the Marchenko method is a costly process in 3D. The retrieval of the homogeneous Green's function for each depth level, in both the inline and the crossline direction, takes 8 hours and 45 minutes on average on 40 Intel E5-2560 cores with a clock speed of 2.3GHz on a node using 256GB of 2133MHz RAM. This computational cost depends on how many virtual source receiver pairs are retrieved and on the hardware that is utilized for the retrieval. While the method is currently costly to implement, recent studies have shown potential for more efficient implementations of the Marchenko method, for example through the use of GPUs \cite{Ravasi2020arXiv} or by employing virtual plane-wave sources instead of virtual point sources \cite{meles2020data}. 

As the reflection data are measured over a medium that is attenuating, the reflection data should be processed in order to compensate for these losses, which has been achieved in practice as well \cite{brackenhoff2019virtual,brackenhoff2019}. It should be noted that the Marchenko method presented in this paper is based on the acoustic wave equation, while the solid Earth is an elastic medium. The Marchenko method is therefore usually employed in marine settings for limited source-receiver offsets, however, there is ongoing research into the elastic application of the Marchenko method and the single-sided representation \cite{reinicke2019}.

Additionally, in order to apply the single-sided representation on field data for the purpose of monitoring the wavefield caused by an actual source, there are some requirements that should be taken into account. First of all, the wavefield caused by the actual source needs to be recorded at monitoring locations that coincide with the receivers of the reflection response. Secondly, the frequency content of the recorded wavefield and the reflection response should overlap, at least partially. The reflection data should ideally be recorded near a monitoring array so that the location requirement is fulfilled. The frequency content of induced events recorded by this array should be studied before the active survey is performed so that the source signal of the active survey can be adjusted accordingly. While these requirements are not strictly necessary for the forecasting approach, utilizing a frequency bandwidth that is similar to that of local induced events would allow for a more realistic forecasting of the wavefield.

\section{Conclusions}
We have shown that the Marchenko method can be applied to 3D reflection data at the surface of the Earth to obtain the responses for virtual source-receiver pairs in the subsurface in a data-driven way. We did this by considering the 3D single-sided representation for obtaining the homogeneous Green's function in the subsurface. This research was previously performed for 2D settings and from a theoretical standpoint, the 3D extension was straightforward, however, the practical implementation placed a strong demand on the quality of the reflection data, especially in the sampling of the data. When this demand was met, we retrieved a homogeneous Green's function that shows a strong match with a reference result. The retrieval of the single-sided causal Green's function showed a similar accuracy when the virtual receiver is located above the virtual source, however, this accuracy decreased significantly when the virtual receiver was located below the virtual source. The results further showed that the quality of the retrieved homogeneous Green's function decreased near the edge of the aperture and with increasing depth. The acquisition geometry therefore determined at which location in the subsurface we could obtain an accurate result.

We also studied the inclusion of 3D rupture mechanisms in the retrieval of the virtual wavefield, through the use of the moment tensor and time-shifting and superposing of point source responses. This was an important extension, as the radiation pattern of a 3D moment tensor source is more complex than that of a 2D version of the same source. The part of the radiation pattern of the source that could be properly retrieved was also limited by the spatial extent and placement of the aperture. When the single-sided representation was used to create both virtual receivers and virtual sources, the wavefield could be predicted in any location in the subsurface that was illuminated by the reflection data. When the wavefield caused by a real source in the subsurface was monitored, issues arose when the source was active over a period of time instead of being impulsive. The accuracy of the result decreased significantly at depths below the source, consistent with the theory. However, the shallow part of the subsurface could be monitored, and the late coda was retrieved accurately for all depths. This information is relevant for studying the effects that a seismic event can have on the shallow subsurface. Further development of the method can yield more accurate results, which in turn can assist in the determination of seismic hazard.

\section*{Copyright statement}
© 2021 IEEE.  Personal use of this material is permitted.  Permission from IEEE must be obtained for all other uses, in any current or future media, including reprinting/republishing this material for advertising or promotional purposes, creating new collective works, for resale or redistribution to servers or lists, or reuse of any copyrighted component of this work in other works.

\section*{Software}
The 3D Marchenko code that was developed for this paper is fully opensource and can be found at \path{https://github.com/JanThorbecke/OpenSource} in the subfolder \texttt{marchenko3D}. The 3D finite-difference modeling code that was used in this paper can be found at the same address in the subfolder \texttt{fdelmodc3D}. The 3D figures in this paper were created using the ParaView software \cite{Ayachit2015}.

\section*{Acknowledgments}
This work has received funding from the European Union's Horizon 2020 research and innovation program: European Research Council (grant agreement no. 742703).\\
The authors wish to thank two anonymous reviewers for their comments and suggestions to improve this paper.

\bibliographystyle{apalike}
\bibliography{VS3D}

\end{document}